\documentclass[runningheads,a4paper]{llncs}

\usepackage{color}
\usepackage{comment}
\usepackage{graphicx}
\usepackage{mathtools}
\usepackage{booktabs}
\usepackage{multirow}
\usepackage[tight]{subfigure}
\usepackage{amsmath, amssymb}
\usepackage{xspace}
\usepackage{nicefrac}

\usepackage{amsthm}
\usepackage{url}
\usepackage{fixfoot}
\urldef{\mailsa}\path|{alfred.hofmann, ursula.barth, ingrid.haas, frank.holzwarth,|
\urldef{\mailsb}\path|anna.kramer, leonie.kunz, christine.reiss, nicole.sator,|
\urldef{\mailsc}\path|erika.siebert-cole, peter.strasser, lncs}@springer.com|    
\newcommand{\keywords}[1]{\par\addvspace\baselineskip
\noindent\keywordname\enspace\ignorespaces#1}
\DeclareFixedFootnote{\rep}{contributed equally to this work}

\begin{document}

\mainmatter  % start of an individual contribution

\title{Multiscale Identification of Topological Domains in Chromatin}
\renewcommand\footnotemark{}
\titlerunning{Multiscale Identification of Topological Domains in Chromatin}

\author{Darya Filippova$^{1,2,\star}$\thanks{*contributed equally to this work} \and Rob Patro$^{2,\star}$ \and Geet Duggal$^{1,2,\star}$ \and Carl Kingsford$^{2}$}%

\institute{1 --- Joint Carnegie Mellon University --- University of Pittsburgh Ph.D. Program in Computational Biology, Pittsburgh, PA\\2 --- Lane Center for Computational Biology, Carnegie Mellon University, Pittsburgh, PA}


%\toctitle{Lecture Notes in Computer Science}
%\tocauthor{Authors' Instructions}
\maketitle

%%%%%%%%%%%%%%%%%%%%%%%%%%%%%%%%%%%%%%%%%%%%%
%
% Abstract
%
%%%%%%%%%%%%%%%%%%%%%%%%%%%%%%%%%%%%%%%%%%%%%
\begin{abstract}

Recent chromosome conformation capture experiments have led to the discovery of dense, contiguous, megabase-sized topological domains that are similar across cell types and conserved across species.  These domains are strongly correlated with a number of chromatin markers and have since been included in a number of analyses. However, functionally-relevant domains may exist at multiple length scales.  We introduce a new and efficient algorithm that is able to capture persistent domains across various resolutions by adjusting a single scale parameter. The identified novel domains are substantially different from domains reported previously and are highly enriched for insulating factor CTCF binding and histone modfications at the boundaries.

\end{abstract}
%%%%%%%%%%%%%%%%%%%%%%%%%%%%%%%%%%%%%%%%%%%%%
%
% Introduction
%
%%%%%%%%%%%%%%%%%%%%%%%%%%%%%%%%%%%%%%%%%%%%%
\section{Introduction}

Chromatin interactions obtained from a variety of recent experimental techniques in chromosome conformation capture (3C)~\cite{DeWit2012} have resulted in significant advances in our understanding of the geometry of chromatin structure~\cite{Gibcus2013}, its relation to the regulation of gene expression, nuclear organization, cancer translocations~\cite{Cavalli2013}, and copy number alterations in cancer~\cite{Fudenberg2011}.  Of these advances, the recent discovery of dense, contiguous regions of chromatin termed \emph{topological domains}~\cite{Dixon2012} has resulted in the incorporation of domains into many subsequent analyses~\cite{Hou2012,Kolbl2012,Lin2012} due to the fact that they are persistent across cell types, conserved across species, and serve as a skeleton for the placement of many functional elements of the genome~\cite{Bickmore2013a,Tanay2013}.
 
\begin{figure}[t!]
	\begin{center}
		\includegraphics[width=3in]{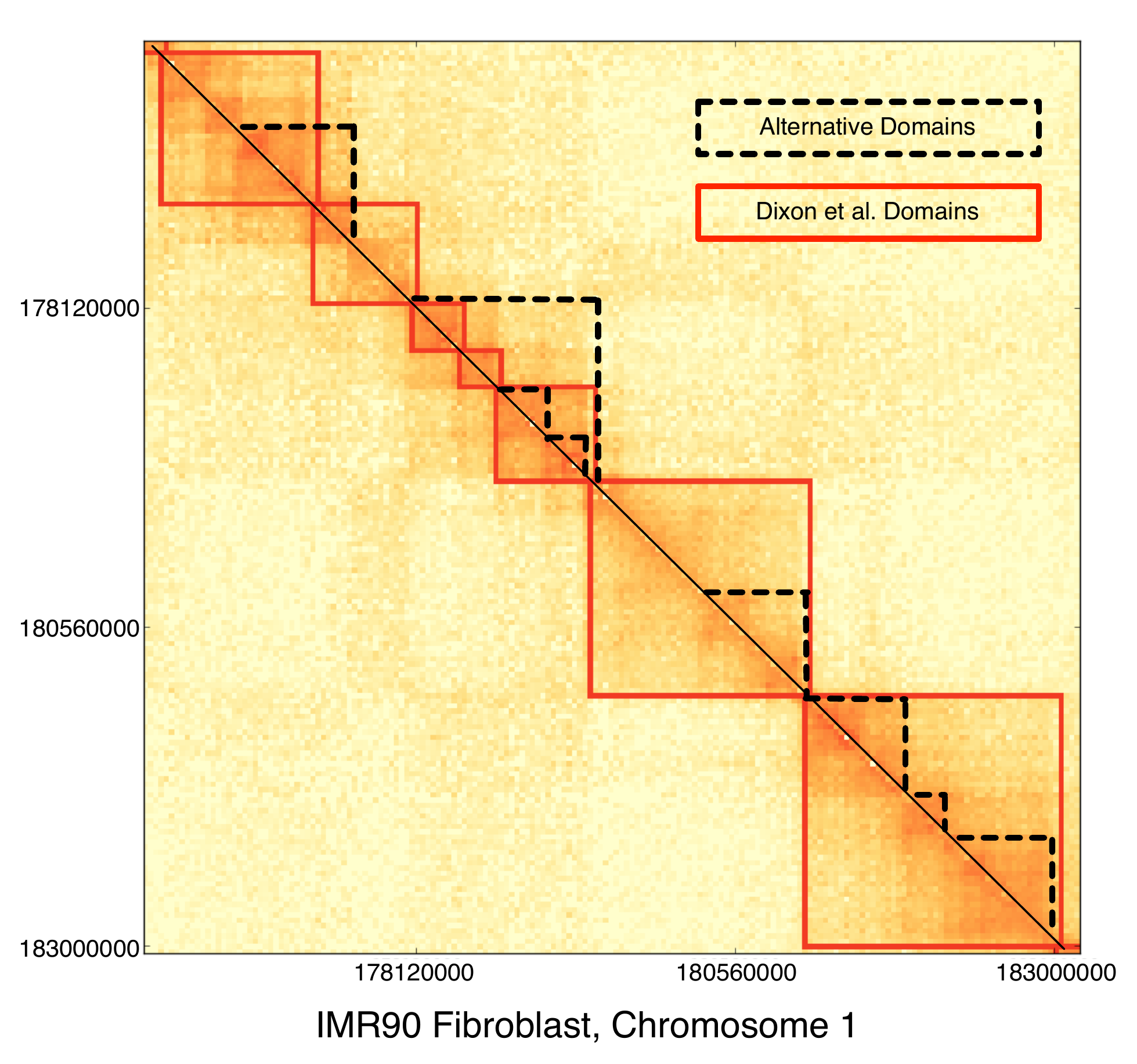}
	\end{center}
	\caption{Interaction matrix for a portion of human chromosome 1 from a recent Hi-C experiment by Dixon et al.~\cite{Dixon2012}. Each axis represents a location on the chromosome (40kbp bins).  Densely interacting domains identified by the method of Dixon et al. (red boxes).  Alternative domains are shown as dotted black lines on the upper triangular portion of the matrix.  Visual inspection of the lower triangular portion suggests domains could be completely nested within another and highly overlapping when compared to Dixon et al.'s domains. This motivates the problem of identifying alternative domains across length scales.}
	\label{heatmap}
\end{figure}

3C experiments result in matrices of counts that represent the frequency of cross-linking between restriction fragments of DNA that are spatially near one another. The original identification of domains in Dixon et al.~\cite{Dixon2012} employed a Hidden Markov Model (HMM) on these interaction matrices to identify regions initiated by significant downstream chromatin interactions and terminated by a sequence of significant upstream interactions.  A defining characteristic of the domains resulting from their analysis is that higher frequency 3C interactions tend to occur within domains as opposed to across domain. This aspect of domains is also reflected in the block-diagonal structure of 3C interaction matrices as shown in Fig.~\ref{heatmap}. In this sense, domains can be interpreted as contiguous genomic regions that self-interact frequently and are more spatially compact than their surrounding regions.

However, the single collection of megabase-sized domains may not be the only topologically and functionally relevant collection of domains. On closer inspection of the block-diagonal matrix structure in Fig.~\ref{heatmap}, it becomes clear that there are alternative contiguous regions of the chromosome that self-interact frequently and are likely more spatially compact than their surrounding regions (dotted lines).  Some of these regions appear to be completely nested within others, suggesting a hierarchy of compact regions along the chromosome, while others appear to overlap each other. These observations suggest that functionally-relevant chromosomal domains may exist at multiple scales.

We introduce a new algorithm to efficiently identify topological domains in 3C interaction matrices for a given domain-length scaling factor $\gamma$. Our results suggest that there exist a handful of characteristic resolutions across which domains are similar. Based on this finding, we identify a consensus set of domains that persist across various resolutions. We find that domains discovered by our algorithm are dense and cover interactions of higher frequency than inter-domain interactions. Additionally, we show that inter-domain regions within the consensus domain set are highly enriched with insulator factor CTCF and histone modification marks. We argue that our straightforward approach retains the essence of the more complex multi-parameter HMM introduced in~\cite{Dixon2012} while allowing for the flexibility to identify biologically relevant domains at various scales.

%%%%%%%%%%%%%%%%%%%%%%%%%%%%%%%%
% heatmap figure with domains
%%%%%%%%%%%%%%%%%%%%%%%%%%%%%%%%

%%%%%%%%%%%%%%%%%%%%%%%%%%%%%%%%%%%%%%%%%%%%%
%
% Problem definition
%
%%%%%%%%%%%%%%%%%%%%%%%%%%%%%%%%%%%%%%%%%%%%%
\section{Problem definitions}
\theoremstyle{definition}
\newtheorem{prob}{Problem}

Given the resolution of the 3C experiment (say, 40kb), the chromosome is broken into $n$ evenly sized fragments. 3C contact maps record interactions between different sections of the chromosome in the form of a weighted adjacency matrix $\mathbf{A}$ where two fragments $i$ and $j$ interact with frequency $\mathbf{A}_{ij}$.

\begin{prob}[Resolution-specific domains] \label{domainprob} Given a $n \times n$ weighted adjacency matrix $\mathbf{A}$ and a resolution parameter $\gamma \geq 0$, we wish to identify a set of domains $D_{\gamma}$ where each domain is represented as an interval $d_i=[a_i, b_i]$, $1 \leq a_i < b_i \leq n$ such that no two $d_i$ and $d_j$ overlap for any $i \ne j$. Additionally, each domain should have a larger interaction frequency within domain than to its surrounding regions. 

Here, the parameter $\gamma$ is inversely related to the average domain size in $D_{\gamma}$: lower $\gamma$ results in sets of larger domains and higher $\gamma$ corresponds to sets of smaller domains. We define $\gamma$ and discuss it in more detail later in the text.

Specifically, we seek to identify a set of non-overlapping domains $D_{\gamma}$ that optimize the following objective:
\begin{align}
	\label{obj}
	\max \sum_{[a_i,b_i] \in D_{\gamma}} q(a_i,b_i,\gamma),
\end{align}
where $q$ is a function that quantifies the quality of a domain $[a_i, b_i]$ at resolution $\gamma$. Since domains are required to contain consecutive fragments of the chromosome, this problem differs from the problem of clustering the graph of 3C interactions induced by $\mathbf{A}$, since such a clustering may place non-contiguous fragments of the chromosome into a single cluster. In fact, this additional requirement allows for an efficient optimal algorithm.
\end{prob}

\begin{prob}[Consensus domains across resolutions]
\label{consensusprob}  Given $\mathbf{A}$ and a
set of resolutions $\Gamma = \{\gamma_1, \gamma_2, \ldots \}$, identify a set of non-overlapping domains $D_c$ that are most persistent across resolutions in $\Gamma$:
\begin{align}
\label{consobj}
\max \sum_{[a_i,b_i] \in D_c} p(a_i,b_i,\Gamma),
\end{align}	
where $p(a_i,b_i,\Gamma)$ is the persistence of domain $[a_i, b_i]$ corresponding to how often it appears across resolutions.
\end{prob}

%%%%%%%%%%%%%%%%%%%%%%%%%%%%%%%%%%%%%%%%%%%%%
%
% Algorithms - give intuition for what we want to find, intuition ofr gamma
%
%%%%%%%%%%%%%%%%%%%%%%%%%%%%%%%%%%%%%%%%%%%%%
\section{Algorithms}

%%%%%%%%%%%%%%%%%%%%%%%%%%%%%%%%%%%%%%%%%%%%%
%
%%%%%%%%%%%%%%%%%%%%%%%%%%%%%%%%%%%%%%%%%%%%%
\subsection{Domain identification at a particular resolution}
\label{singleres}
Since each row and corresponding column in a 3C interaction matrix encodes a genomic position on the chromosome, we can write the solution to objective~(\ref{obj}) as a dynamic program:
\begin{align}
	\label{pbd}
	\textsf{OPT}_1(l) = \max_{k<l}\{\textsf{OPT}_1(k-1) + \max\{q(k,l,\gamma),0\}\},
\end{align}
where $\textsf{OPT}_1(l)$ is the optimal solution for objective~(\ref{obj}) for the sub-matrix defined by the first $l$ positions on the chromosome ($\textsf{OPT}_1(0) = 0$). The choice of $k$ encodes the size of the domain immediately preceding location $l$. We define negative-scoring domains as non-domains and, as such, only domains with $q > 0$ in the max term in~(\ref{pbd}) are retained.

Our quality function $q$ is:
\begin{align}
	\label{quality} q(k,l,\gamma) &= s(k,l,\gamma)-\mu_s(l-k),\mbox{ where}\\
	\label{sumtri} s(k,l,\gamma) &= \frac{\sum_{g=k}^l \sum_{h=g+1}^l A_{gh}}{(l-k)^\gamma}
\end{align}
is a \emph{scaled density} of the subgraph induced by the interactions $A_{gh}$ between genomic loci $k$ and $l$. Equation~(\ref{quality}) is the zero-centered sum of~(\ref{sumtri}), which is the upper-triangular portion of the submatrix defined by the domain in the interval $[k,l]$ divided by the scaled length $(l-k)^{\gamma}$ of the domain. When $\gamma=1$,  the scaled density is the weighted subgraph density~\cite{goldberg1984finding} for the subgraph induced by the fragments between $k$ and $l$. When $\gamma=2$, the scaled density is half the internal density of a graph cluster~\cite{Schaeffer2007}. For larger values of $\gamma$, the length of a domain in the denominator is amplified, hence, smaller domains would produce larger objective values than bigger domains with similar interaction frequencies. $\mu_s(l-k)$ is the mean value of~(\ref{sumtri}) over all sub-matrices of length $l-k$ along the diagonal of $\mathbf{A}$, and can it be pre-computed for a given $\mathbf{A}$. We disallow  domains where there are fewer than 100 sub-matrices available to compute the mean. By doing this, we are only excluding domains of size larger than $n-100$ fragments, which in practice means that we are disallowing domains that are  hundreds of megabases long.  Values for the numerator in (\ref{sumtri}) are also pre-computed using an efficient algorithm~\cite{Filippova2012}, resulting in an overall run-time of $O(n^2)$ to compute $\textsf{OPT}_1(n)$.

%%%%%%%%%%%%%%%%%%%%%%%%%%%%%%%%%%%%%%%%%%%%%
% Consensus set - persistence
%%%%%%%%%%%%%%%%%%%%%%%%%%%%%%%%%%%%%%%%%%%%%
\subsection{Obtaining a consensus set of persistent domains across resolutions}
\label{consensusalg}
For objective (\ref{consobj}), we use the procedure in section~\ref{singleres} to construct a set $\mathcal{D} = \bigcup_{\gamma \in \Gamma} D_{\gamma}$.  $\mathcal{D}$ is a set of overlapping intervals or domains, each with a quality score defined by its persistence $p$ across resolutions. To extract a set of highly persistent, non-overlapping domains from $\mathcal{D}$, we reduce  problem~\ref{consensusprob} to the weighted interval scheduling problem~\cite{Kleinberg2005}, where competing requests to reserve a resource in time are resolved by finding the highest-priority set of non-conflicting requests. To find a consensus set of domains, we map a request associated with an interval of time to a domain and its corresponding interval on the chromosome. The priority of a request maps to a domain's persistence $p$ across length scales.

The algorithm to solve problem~\ref{consensusprob} is then:
\begin{align}
\label{wis}
\textsf{OPT}_2(j) = \max\{\textsf{OPT}_2(j-1), \textsf{OPT}_2(c(j)) + p(a_j,b_j,\Gamma) \}
\end{align}
where $\textsf{OPT}_2(j)$ is the optimal non-overlapping set of domains for the $j$th domain in a list of domains sorted by their endpoints ($\textsf{OPT}_2(0) = 0$), and $c(j)$ is the closest domain before $j$ that does not overlap with $j$.  The first and second terms in~(\ref{wis}) correspond to either choosing or not choosing domain $j$ respectively.
We pre-compute a domain's persistence $p$ as:
\begin{align}
\label{persist}
p(a_i,b_i,\Gamma) = \sum_{\gamma \in \Gamma} \delta_i \text{ where }
\delta_i = \begin{cases}
1 & \text{if } [a_i,b_i] \in D_{\gamma} \\ 0 & \text{otherwise.}
\end{cases}
\end{align}
Equation~(\ref{persist}) is therefore a count of how often domain $i$ appears across all resolutions in $\Gamma$ for domain sets identified by the method in section~\ref{singleres}. It may be desirable to treat multiple highly overlapping, non-equivalent domains as a single domain, however, we conservatively identify exact repetitions of a domain across resolutions since this setting serves as a lower bound on the persistence of the domain. If $m=|\mathcal{D}|$, then pre-computing persistence takes $O(m|\Gamma|)$ time, and $c(j)$ is precomputed after sorting the intervals by their endpoints. The limiting factor when computing $\textsf{OPT}_2(m)$ is time to compute $c(j)$, which is $m\log m$. Thus, the overall algorithm runs in $O(m\log m + (n^2+m)|\Gamma|)$ time taking into account an additional $O(n^2|\Gamma|)$ for computing $\mathcal{D}$.

%%%%%%%%%%%%%%%%%%%%%%%%%%%%%%%%%%%%%%%%%%%%%
%
% Results: intrinsic and extrinsic validation of domains, comparison 
% to Bing Ren's domains
%
%%%%%%%%%%%%%%%%%%%%%%%%%%%%%%%%%%%%%%%%%%%%%
\section{Results}

We used chromatin conformation capture data from Dixon et al.~\cite{Dixon2012} for human fibroblast and mouse embryonic cells. The 3C contact matrices were already aggregated at fragment size 40kb and were corrected for experimental bias according to~\cite{Yaffe2011}. We compared our multiscale domains and consensus sets against the domains generated by Dixon et al. for the corresponding cell type and species. 
For human fibroblast cells, we used CTCF binding sites from~\cite{Kim2007}.
For mouse embryonic cell CTCF binding sites and chromatin modification marks, we used data by Shen et al.~\cite{Shen2012}.

% List of parameters required by Dixon et al.
% \begin{itemize}
% \item 2MB upstream/downstream boundary
% \item 1-20 mixtures of gaussians
% \item Median posterior probabilities $\geq 0.99$, at least 80kbp
% \end{itemize}

%%%%%%%%%%%%%%%%%%%%%%%%%%%%%%%%%%%%%%%%%%%%%
% Dense domains
%%%%%%%%%%%%%%%%%%%%%%%%%%%%%%%%%%%%%%%%%%%%%
\subsection{Ability to identify densely interacting domains across scales}

%%%%%%%%%%%%%%%%%%%%%%%%%%%%%%%%%%%%
% inter-intra dist plot; disributions of sizes
%%%%%%%%%%%%%%%%%%%%%%%%%%%%%%%%%%%%
\begin{figure}[t]
	\begin{center}
	\subfigure[Domain size vs. frequency]{
		\includegraphics[width=0.45\linewidth]{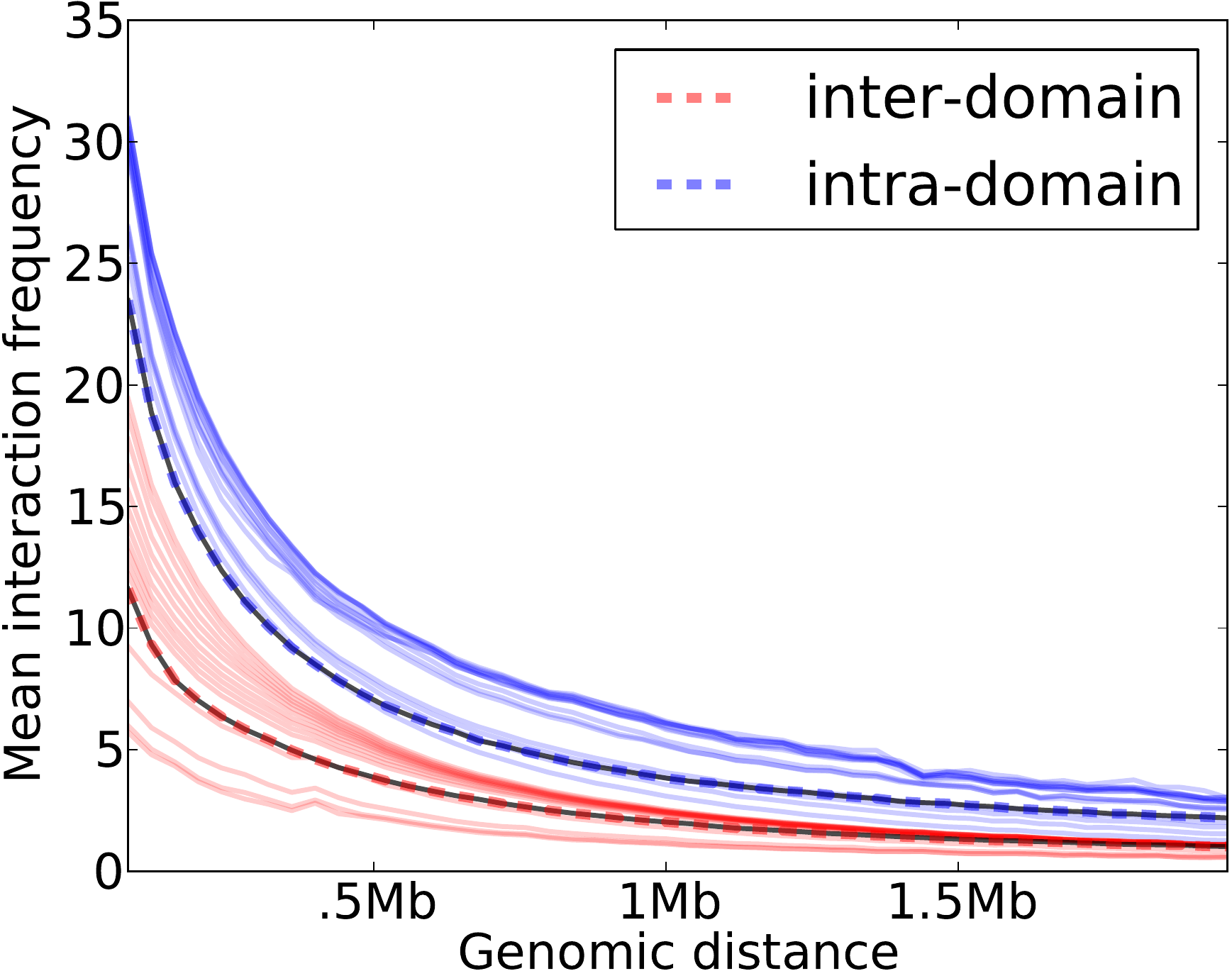}
		\label{subfig:size-freq}
	}
	\subfigure[Mean frequency distr.]{
		\includegraphics[width=0.45\linewidth]{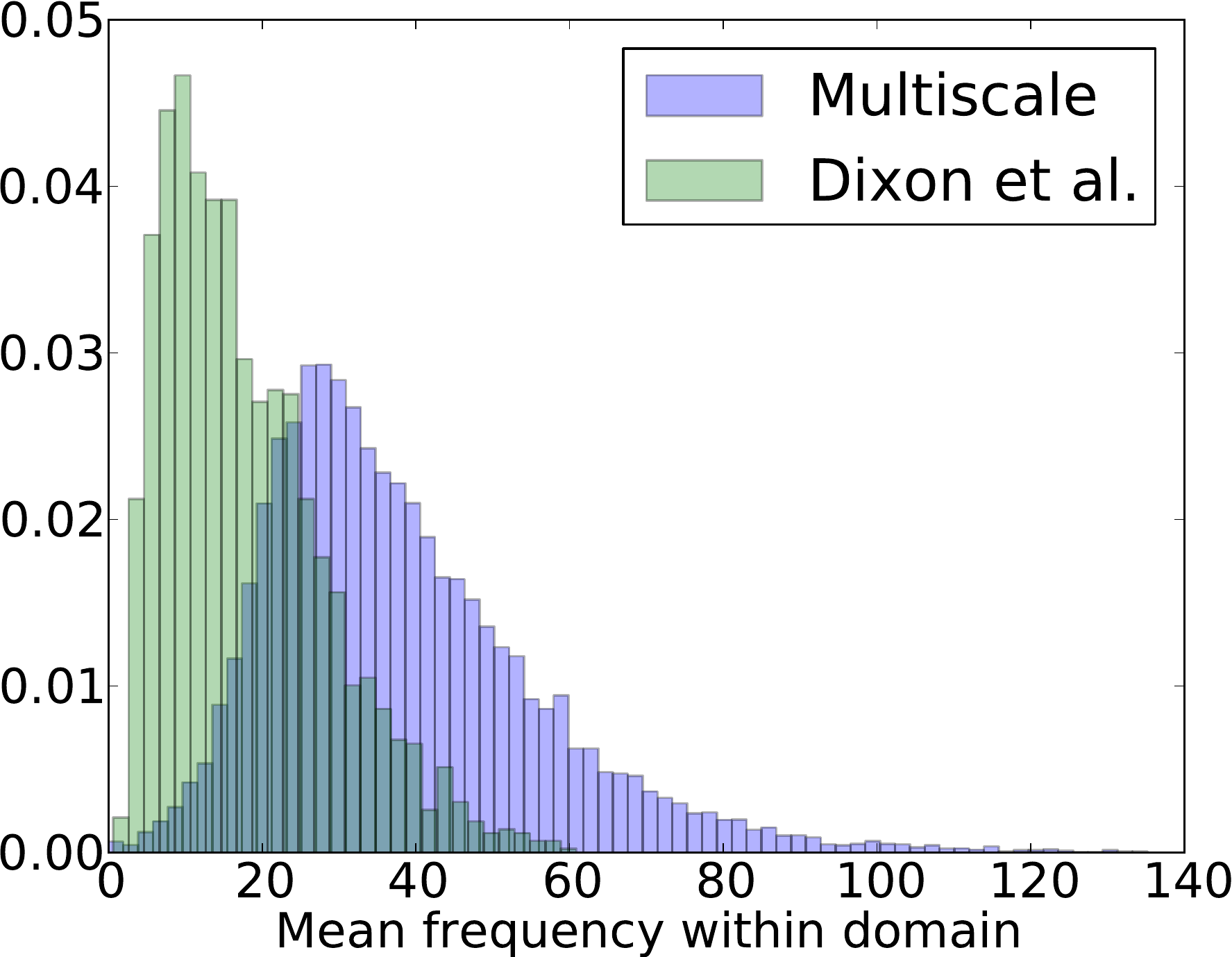}
		\label{subfig:freq-distr}
	}
	\caption{~\subref{subfig:size-freq} Our algorithm discovers domains with mean frequency value for inter- and intra-domain interactions (solid lines) at or better than that of Dixon et al. domains (dotted lines). Each solid line represents domains at different resolution $\gamma$ in human fibroblast cells. \subref{subfig:freq-distr}~Multiscale domains identified in human fibroblast cells by our dynamic program tend to have higher mean frequency than those of Dixon et al.
	(distributions are plotted after outliers $> \mu+4\sigma$ were removed).}
	
	\label{fig:mi_jacc}
	\end{center}
\end{figure}

Multiresolution domains successfully capture high frequency interactions and leave interactions of lower mean frequency outside of the domains. We compute the mean interaction frequency for all intra- and inter-domain interactions at various genomic lengths and plot the distribution of means for multiple resolutions (Fig.~\ref{subfig:size-freq}). The mean intra-domain interaction frequency (blue) is consistently higher (up to two times) than the mean frequency for interactions that cross domains (red). Compared to the domains reported by Dixon et al., our domains tend to aggregate interactions of higher mean frequency, especially at larger $\gamma$. The distribution of mean intra-domain frequencies for Dixon et al. is skewed more to the left than that of the multiscale domains (Fig.~\ref{subfig:freq-distr}). This difference can be partially explained by the fact that multiscale domains on average are smaller in size ($\mu=0.2$Mb, $\sigma=1.2$Mb) than domains reported by Dixon et al. ($\mu=1.2$Mb, $\sigma=0.9$Mb).

%%%%%%%%%%%%%%%%%%%%%%%%%%%%%%%%%%%%%%%%%%%%%
%
%%%%%%%%%%%%%%%%%%%%%%%%%%%%%%%%%%%%%%%%%%%%%
\subsection{Domain persistence across scales}

\begin{figure}[t]
	\centering
	\subfigure[domain size \& count vs. $\gamma$]{
  	  \includegraphics[width=0.46\linewidth]{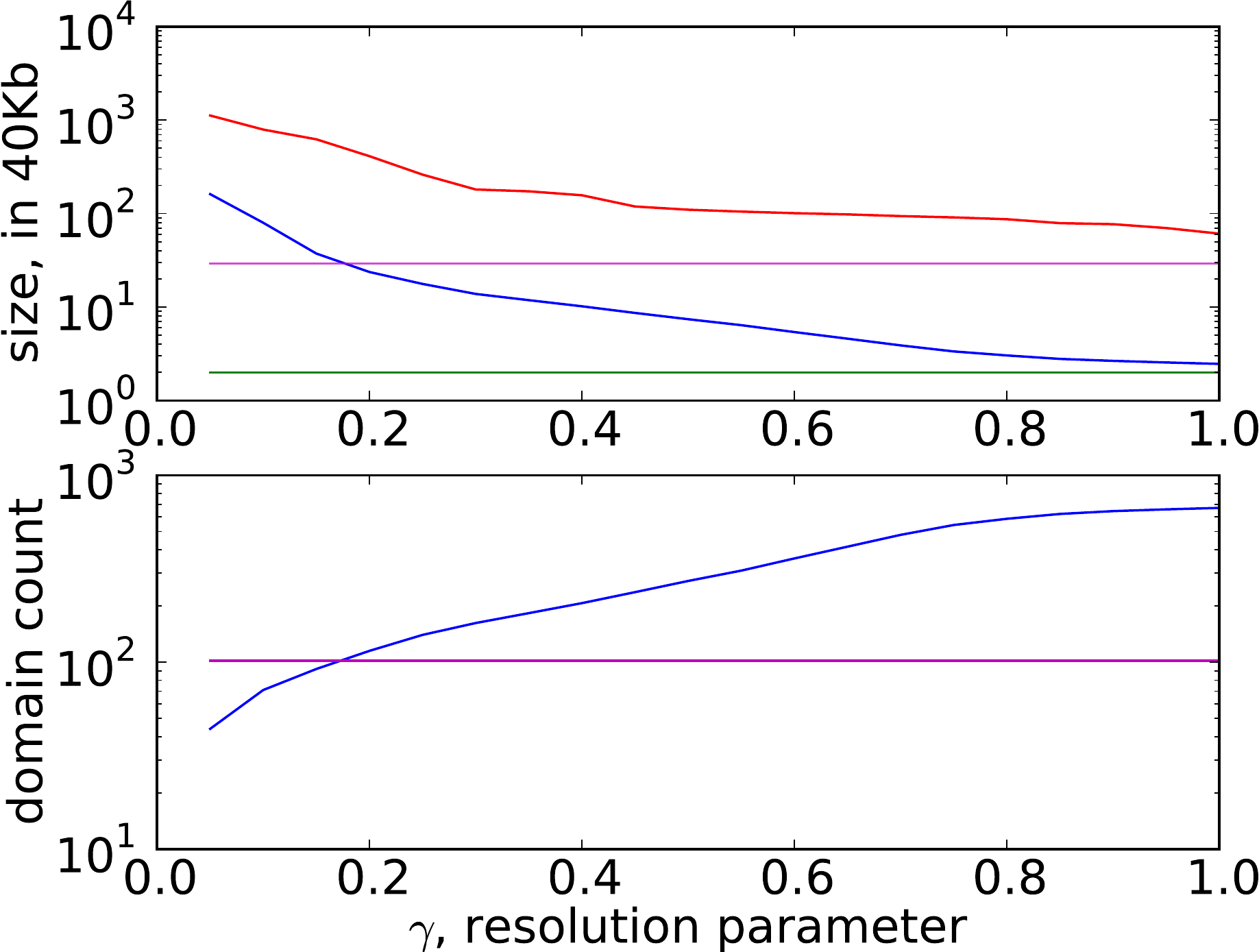}
  	  \label{subfig:sizeCount}
	}
	\subfigure[similarity to Dixon et al. domains]{
	  \includegraphics[width=0.43\linewidth]{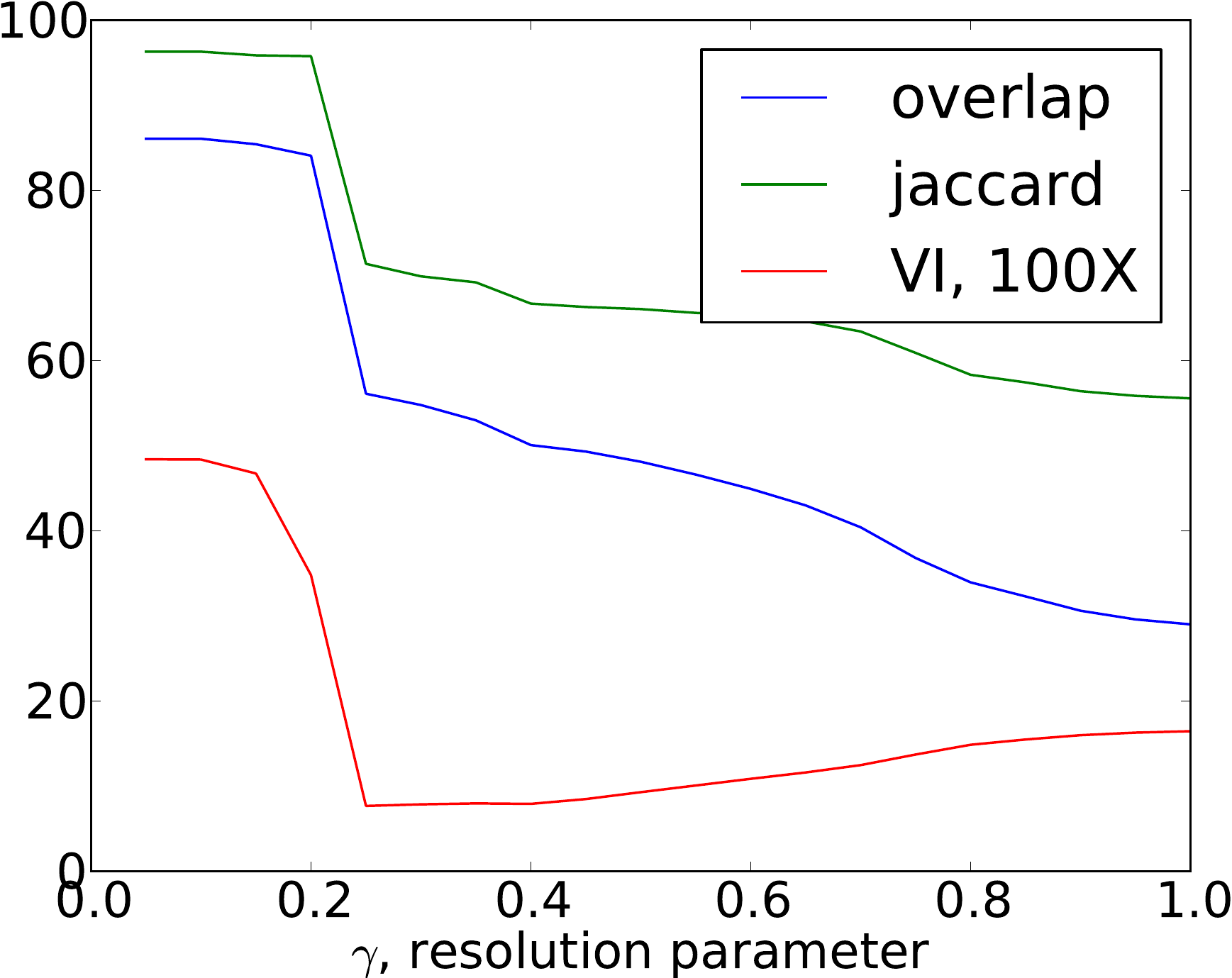}
	  \label{subfig:dixonSim}
	}
	\caption{~\subref{subfig:sizeCount} The domain sizes increase and the domain count decreases as the resolution parameter drops. Above: plotted are maximum (red), average (blue), and minimum (green) domain size averaged over all chromosomes for the domains on human fibroblasts. The magenta line shows the average domain size for domains reported by Dixon et al. Below: the number of domains increases for higher values of resolution parameter. The magenta line displays domain count for Dixon et al.~\subref{subfig:dixonSim} According to the Jaccard metric, the similarity between multiresolution domains and domains reported by Dixon et al. increases as the resolution parameter goes to zero.}%, however, the variation of information suggests that the two sets of domains are most similar around $\gamma=0.25$.}
	\label{fig:dom_size}
\end{figure}

\begin{figure}[t]
	\begin{center}
	\subfigure[domains across resolutions]{
		\includegraphics[width=0.55\linewidth]{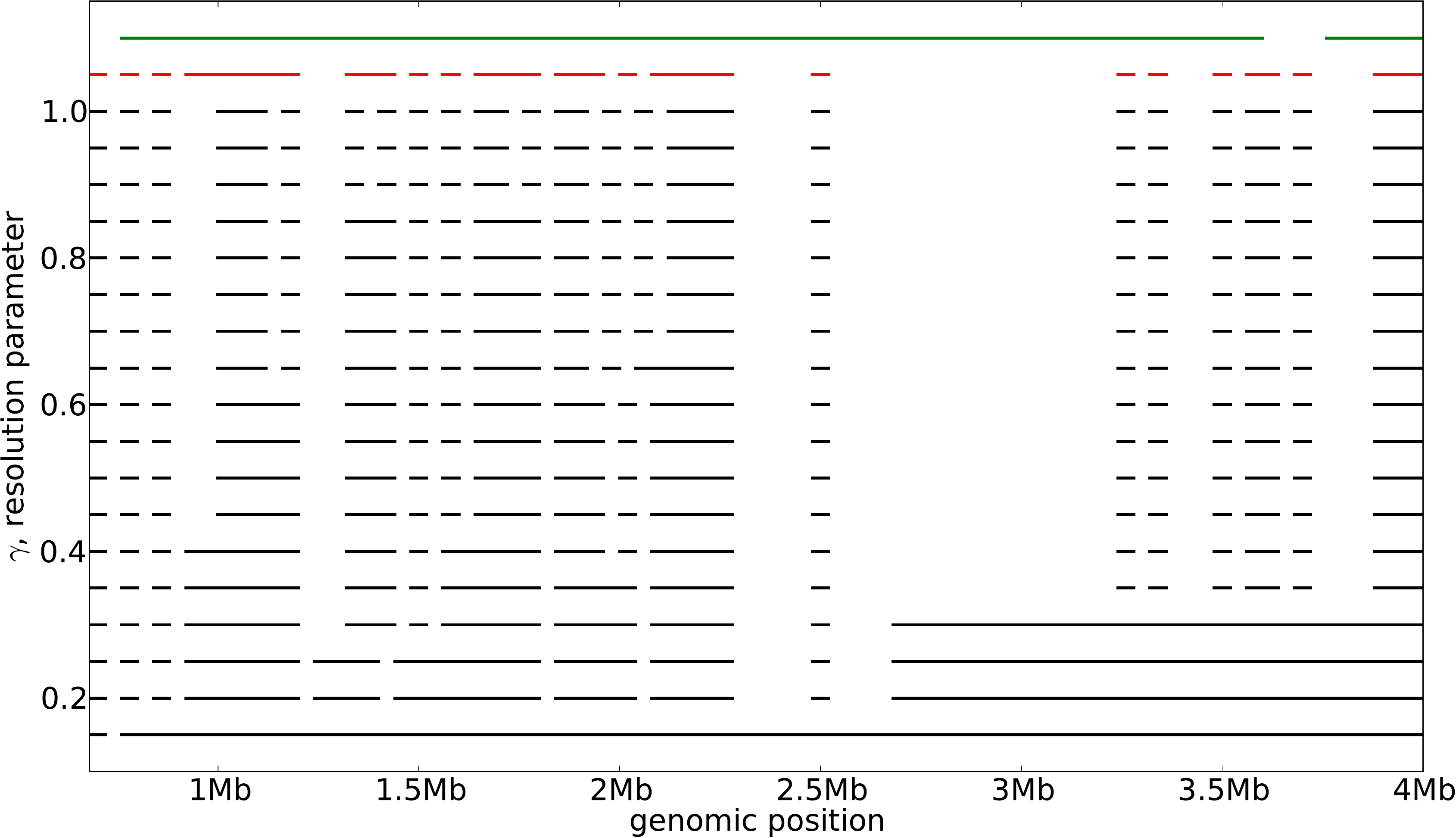}
		\label{subfig:multiresDomains}
	}
	\subfigure[VI across resolutions]{
		\includegraphics[width=0.366\linewidth]{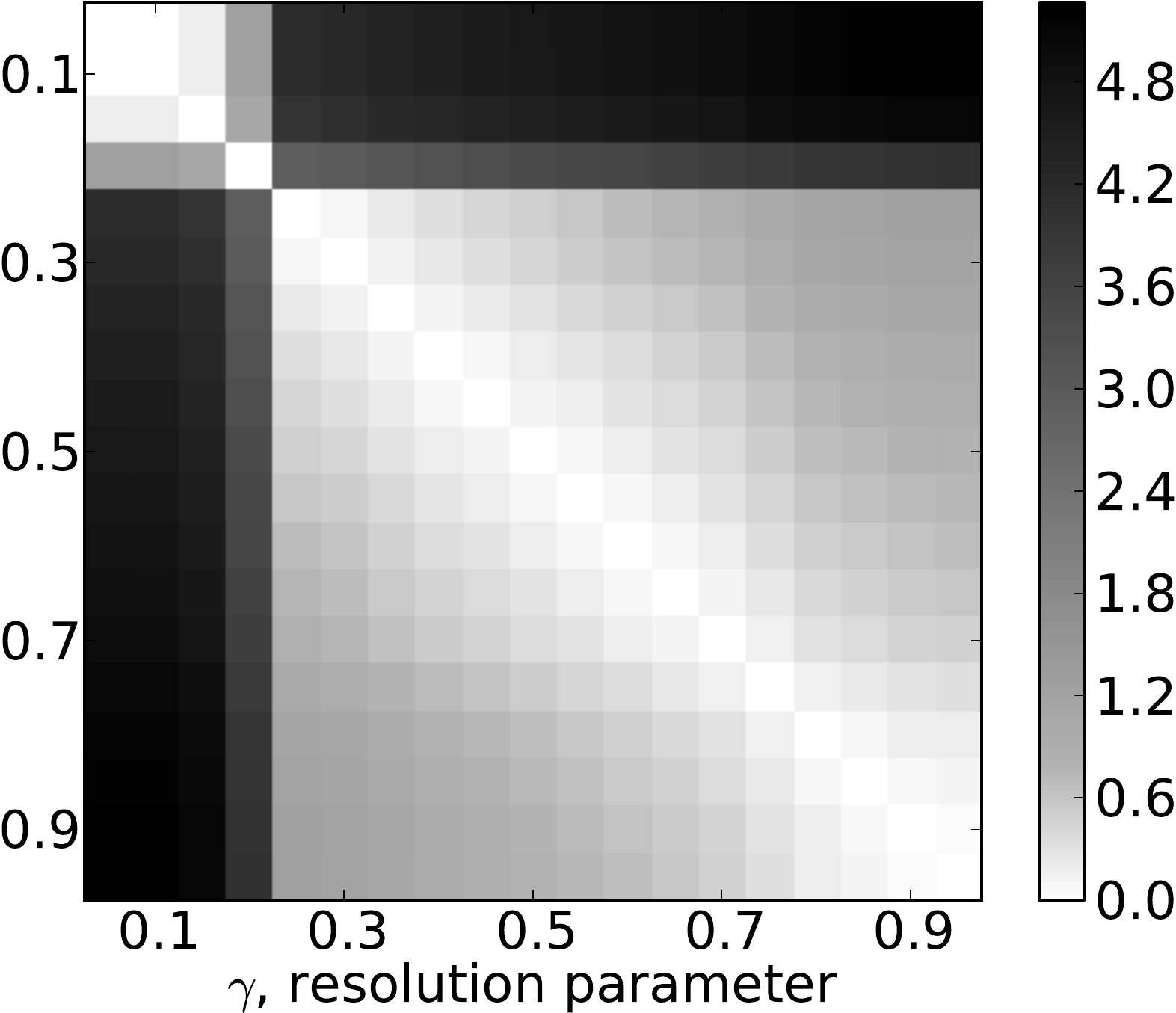}
		\label{subfig:multiresVI}
	}
	\end{center}
	\caption{~\subref{subfig:multiresDomains} Domains identified by our algorithm (black) are smaller at higher resolutions and merge to form larger domains at $\gamma$ close to 0. Visual inspection shows qualitative differences between consensus domains (red) and domains reported by Dixon et al. (green). Data shown for the first 4Mb of chromosome 1.~\subref{subfig:multiresVI} Variation of information for domains identified by our algorithm across different resolutions for chromosome 1 in human fibroblast cells.}
	\label{fig:domains_line}
\end{figure}

Domain sets across resolutions share significant similarities, even as the distribution of domains and their sizes begin to change (Fig.~\ref{fig:dom_size}). The patterns of similarity are particularly obvious if we plot the domains at various resolutions (Fig.~\ref{subfig:multiresDomains}): many domains identified by our algorithm persist at several resolutions and are aggregated into larger domains at smaller $\gamma$, suggesting a hierarchical domain structure. The stability of these domains across resolutions indicates that the underlying chromosomal structure is dense within these domains and that these domains interact with the rest of the chromosome at a much lower frequency.

%%%%%%%%%%%%%%%%%%%%%%%%%%%%%%%%%%%%%%%%%%%%
% Domain sizes over all resolutions, jaccard/overlap/vi between us and b.r. 
% over all resolutions
%%%%%%%%%%%%%%%%%%%%%%%%%%%%%%%%%%%%%%%%%%%%

A pairwise comparison of domain configurations displays regions of stability across multiple resolutions (Fig.~\ref{subfig:multiresVI}). We use the variation of information (VI)~\cite{Meila2003}, a metric for comparing two sets of clusters, to compute the distance between two sets of domains. To capture the similarities between two domain sets $D$ and $D'$ and the inter-domain regions induced by the domains, we construct new derivate sets $C$ and $C'$ where $C$ contains all domains $d \in D$ as well as all inter-domain regions ($C'$ is computed similarly). To compute entropy $H(C) = \sum_{c_i  \in C} p_i \log p_i$, we define the probability of seeing each interval in $C$ as $p_i = (b_i - a_i) / L$ where $L$ is the number of nucleotides from the start of the leftmost domain to the end of the rightmost domain in the set $D \cup D'$. When computing the mutual information $I(C, C') = \sum_{c_i \in C} \sum_{c'_j \in C'} p_{ij} \log[ p_{ij} / (p_i p_j) ]$ between two sets of intervals $C$ and $C'$, we define the joint probability $p_{ij}$ to be $| [a_i, b_i] \cap [a_j, b_j] | / L$.
We then compute variation of information on these two new sets: $VI(C, C') =  H(C) + H(C') - 2I(C, C')$ where $H(\cdot)$ is entropy and $I(\cdot, \cdot)$ is mutual information. Chromosome 1, for example, has three visually pronounced groups of resolutions within which domain sets tend to be more similar than across ($\gamma = $[0.00-0.20], [0.25-0.70], and [0.75-1.00] --- see Fig.~\ref{subfig:multiresVI}).

%%%%%%%%%%%%%%%%%%%%%%%%%%%%%%%%%%%%%%%%%%%%
% Domains at different resolutions as a line plot
%%%%%%%%%%%%%%%%%%%%%%%%%%%%%%%%%%%%%%%%%%%%

\subsection{Comparison with the previously identified set of domains in Dixon et al.}

\begin{figure}[t!]
	\begin{center}
		\includegraphics[width=.9\linewidth]{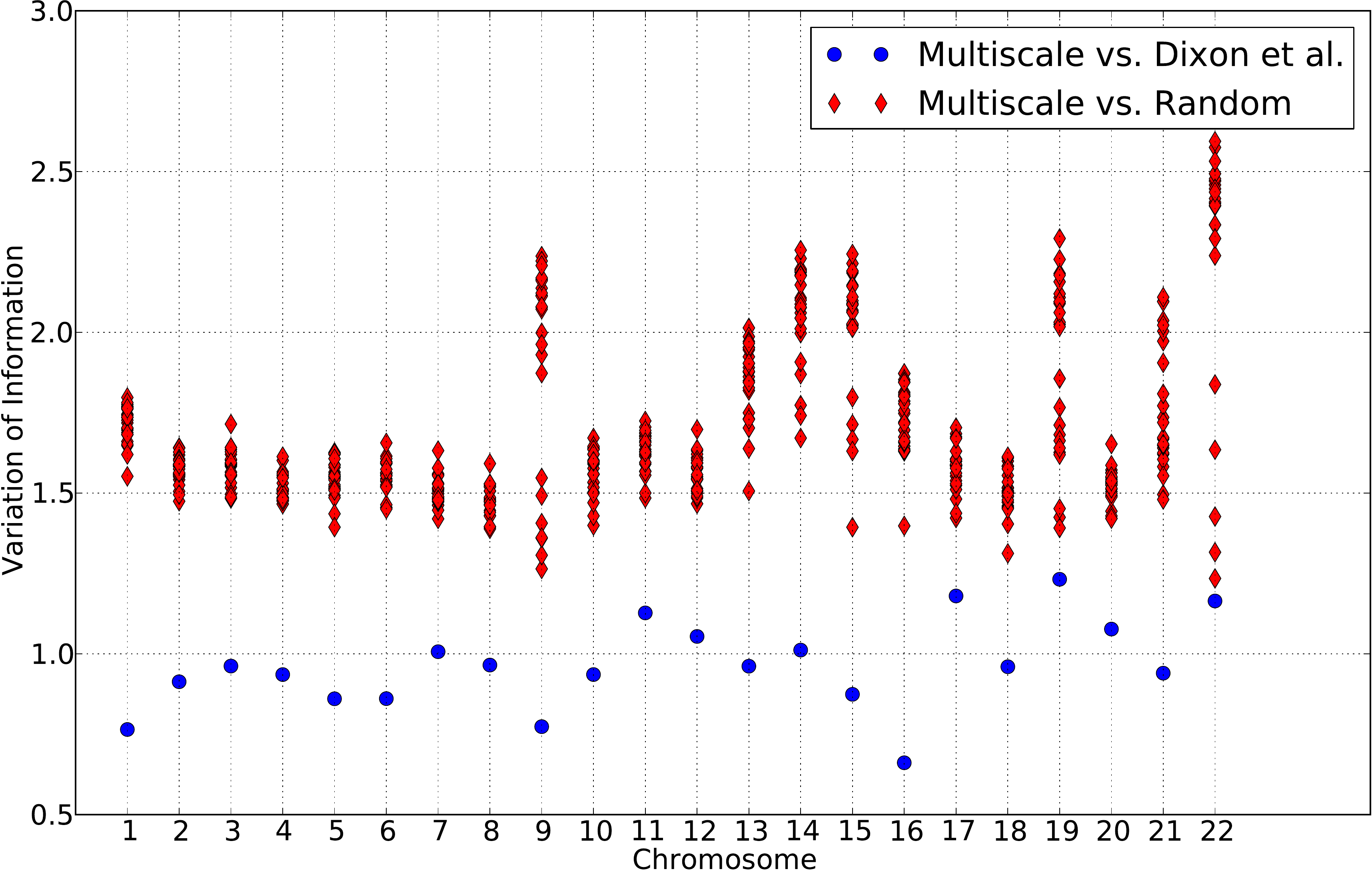}
	\end{center}
	\caption{Comparison of Dixon et al.'s domain set with the multiscale consensus set for chromosomes 1--22 ($x$-axis). We used the variation of information (VI) ($y$-axis) to compute distances between domain sets for the multiscale consensus set vs. Dixon et al (blue dots) and the multiscale consensus vs. randomly shuffled domains (red diamonds).}
	\label{fig:consensus_agreement}
\end{figure}

At higher resolutions, domains identified by our algorithm are smaller than those reported by Dixon et al. (Fig.~\ref{subfig:sizeCount}). As the resolution parameter decreases to 0.0, the average size of the domains increases  (see Fig.~\ref{fig:dom_size} for results for chromosome 1 on the IMR90 human fibroblast cells). As domains expand to cover more and more of the chromosome, the similarity to the domains identified by Dixon et al.~\cite{Dixon2012} also increases (Fig.~\ref{subfig:dixonSim}). We calculate the Jaccard similarity between two sets of domains $D$ and $D'$ as $J(D, D') = N_{11} / (N_{11} + N_{01} + N_{10})$ where the quantities $N_{11}$, $N_{01}$, and $N_{10}$ are the number of 3C fragments that are in a domain in both sets $D$ and $D'$, the number of fragments that are in a domain in $D'$, but not in $D$, and the number of fragments that are in a domain in $D$, but not $D'$, respectively (light blue in Fig.~\ref{subfig:dixonSim}). The composition of the domains, however, is different as is captured by the variation of information (red in Fig.~\ref{subfig:dixonSim}). Overall, we identify domains that cover similar regions of the chromosome (Fig.~\ref{fig:mi_jacc}), yet differ in their size distribution and genomic positions.

We use the algorithm described in section~\ref{consensusalg} to obtain a consensus set of domains $D_c$ persistent across resolutions. We construct the set $\Gamma$ by defining the range of our scale parameter to be $[0, \gamma_\textrm{max}]$ and incrementing $\gamma$ in steps of 0.05. In order to more directly compare with previous results, we set $\gamma_{\max}=0.5$ for human and $0.25$ for mouse since these are the scales at which the maximum domain sizes in Dixon et al.'s sets match the maximum domain sizes in our sets.

Our consensus domain set agrees with the Dixon et al. domains better than with a randomized set of domains adhering to the same domain and non-domain length distributions (Fig.~\ref{fig:consensus_agreement}). Our primary motivation in comparing to randomized sets of domains is to provide a baseline that we can use to contrast our set of domains with Dixon et al. Comparing to a set of random domains also helps to verify that our observations are due to the observed sequence of domains and not the distribution of domain lengths. To shuffle Dixon's domains, we record the length of every domain and non-domain region, and then shuffle these lengths to obtain a randomized order of domains and non-domains across the chromosome.  The fact that variation of information is lower between consensus domains and domains reported by Dixon et al. demonstrates that, though the approaches find substantially different sets of topological domains, they still agree significantly more than one would expect by chance. 

%%%%%%%%%%%%%%%%%%%%%%%%%%%%%%%%%%%%%%%%%%%%%
%
%%%%%%%%%%%%%%%%%%%%%%%%%%%%%%%%%%%%%%%%%%%%%
\subsection{Enrichment of CTCF and histone modifications near boundaries}
\label{sec:Enrichment}

We assess the enrichment of transcription factor CTCF and histone modifications H3K4me3 and H3K27AC within the inter-domain regions induced by the consensus domains. These enrichments provide evidence that the boundary regions between topological domains correlate with genomic regions that act as insulators and barriers, suggesting that the topological domains may play a role in controlling transcription in mammalian genomes~\cite{Dixon2012}.

\begin{figure}
\begin{center}
\includegraphics[width=0.73\textwidth]{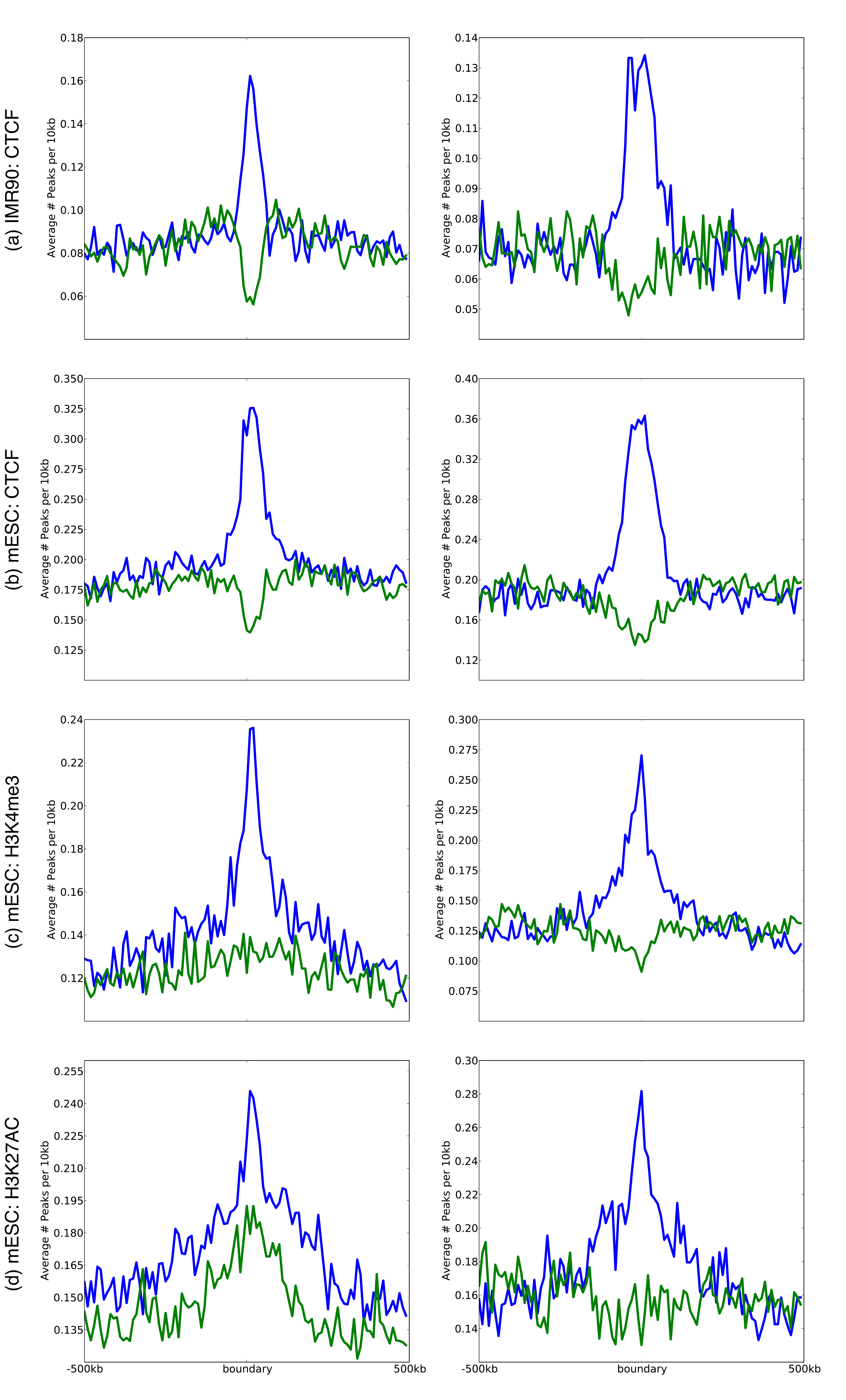}
% \subfigure[IMR90: CTCF]{
% \includegraphics[width=2.1in]{IMR90_CTCF_Ours}
% \includegraphics[width=2.1in]{IMR90_CTCF_Bingy}
% \label{fig:CTCF}
% }
% %
% \subfigure[mESC: CTCF]{
% \includegraphics[width=2.1in]{mESC_CTCF_Ours}
% \includegraphics[width=2.1in]{mESC_CTCF_Bingy}
% \label{fig:mESC_CTCF}
% }
% %
% \subfigure[mESC: H3K4me3]{
% \includegraphics[width=2.1in]{mESC_H3K4me3_Ours}
% \includegraphics[width=2.1in]{mESC_H3K4me3_Bingy}
% \label{fig:H3K4me3}
% }
% %
% \subfigure[mESC: H3K27AC]{
% \includegraphics[width=2.1in]{mESC_H3K27AC_Ours}
% \includegraphics[width=2.1in]{mESC_H3K27AC_Bingy}
% \label{fig:H3K27AC}
% }
\caption{Enrichment of binding CTCF binding (a) in IMR90 and (b) in mESC and histone modifications (c), (d) in mESC around domain boundaries for our consensus set of persistent domains (left, blue), and for those identified by Dixon et al. (right, blue).  Green lines represent the presence of CTCF at the midpoint of the topological domains.}
\label{fig:enrichment}
\end{center}
\end{figure}

Figure~\ref{fig:enrichment} illustrates the enrichment of insulator or barrier-like elements in domain boundaries in both the human fibroblast (IMR90) and mouse embryonic stem cell (mESC) lines.  Specifically, we observe that 
the boundaries between consensus domains are significantly enriched for all of the transcription factors and histone marks we consider.  In certain cases --- specifically in the case of CTCF --- we notice that the CTCF binding signals peak more sharply in the boundaries between the domains we discover than in the boundaries between the domains of Dixon et al.

\begin{table}[b]
\centering
\caption{Each table entry is of the form $\frac{e}{t} \approx r$ where $e$ is the number of elements containing $\ge 1$ of CTCF and histone modifications, $t$ is the total number of elements and $r$ is the approximate ratio $e/t$.  Our method produces more domains,
and hence more boundaries, than that of Dixon et al.~\cite{Dixon2012}.  However, relative to Dixon et al., our domains are depleted for peaks of interest, while our boundaries are significantly enriched
for such peaks.}
%Our method tends to produce ($1.6$---$2.3$ times) more domains than that of Dixon et al.~\cite{Dixon2012}.  However, while the domains produced by both methods contain at least peak for the different chromatin factors we consider in roughly the same proportion, the boundaries between our domains contain at least one peak for these factors about twice as frequently as the boundaries between the domains of Dixon et al.}
\label{tab:differentalEnrichment}
\scriptsize
\begin{tabular}{lc@{\hskip 15pt}c@{\hskip 5pt}|@{\hskip 5pt}c@{\hskip 15pt}c@{\hskip 15pt}c}
\toprule
%& \multicolumn{3}{r}{Ratio: (current method / \cite{Dixon2012})} \\
%\cmidrule(r){2-4}
Signal & Domains (\cite{Dixon2012}) & Domains (Ours) & Boundaries (\cite{Dixon2012}) & Boundaries (Ours) \\
%       & Domains w/ $\ge 1$ peak        & w/ $\ge 1$ peak & w/ $\ge 1$ peak \\
\midrule
CTCF (IMR90)   & $\frac{2050}{2234}\approx0.92$ & $\frac{3092}{5365}\approx0.58$ & $\frac{423}{2136}\approx0.20$ & $\frac{2126}{4861}\approx0.44$ \\[0.5em]
CTCF (mESC)    & $\frac{2057}{2066}\approx1.00$ & $\frac{2500}{3578}\approx0.70$ & $\frac{654}{2006}\approx0.33$ & $\frac{2258}{3122}\approx0.72$ \\[0.5em]
H3K4me3 (mESC) & $\frac{2019}{2066}\approx0.98$ & $\frac{2362}{3578}\approx0.66$ & $\frac{600}{2006}\approx0.30$ & $\frac{1738}{3122}\approx0.60$ \\[0.5em]
H3K27AC (mESC) & $\frac{1922}{2066}\approx0.93$ & $\frac{2254}{3578}\approx0.63$ & $\frac{458}{2006}\approx0.23$ & $\frac{1342}{3122}\approx0.43$ \\
\bottomrule
\end{tabular}
\end{table}
% \vspace{-25px}

We also observe that, when compared with the domain boundaries predicted by Dixon et al., our boundaries more often contain insulator or barrier-like elements (see Table~\ref{tab:differentalEnrichment}). Specifically, we normalize for the fact that we identify approximately twice as many domains as Dixon et al., and generally observe a two-fold enrichment in the fraction of boundaries containing
peaks for CTCF markers. This suggests that structural boundaries identifed by our method are more closely tied to functional sites which serve as barriers to long-range regulation. We also observe a depletion of insulator CTCF elements within our domains when compared to the domains of Dixon et al.  This observation is consistent with the assumption that transcriptional regulation is more active within spatially proximate domains since there are fewer elements blocking regulation within these domains.  Table~\ref{tab:differentalEnrichment} also shows similar patterns for histone modifications which suggests that our domain boundaries are enriched for functional markers of gene regulation.

%%%%%%%%%%%%%%%%%%%%%%%%%%%%%%%%%%%%%%%%%%%%%
%
% Conclusions
%
%%%%%%%%%%%%%%%%%%%%%%%%%%%%%%%%%%%%%%%%%%%%%
\section{Discussion and Conclusions}

In this paper, we introduce an algorithm to identify topological domains in chromatin using interaction matrices from recent high-throughput chromosome conformation capture experiments.  Our algorithm produces domains that display much higher interaction frequencies within the domains than in-between domains (Fig.~\ref{fig:mi_jacc}) and for which the boundaries between these domains exhibit substantial enrichment for several known insulator and barrier-like elements (Fig.~\ref{fig:enrichment}).  To identify these domains, we use a multiscale approach which finds domains at various size scales.  %To obtain a single set of domains from this rich ensemble, e extract a non-overlapping set of consensus domains that are most persistent across multiple length scales.
We define a consensus set to be a set of domains that persist across multiple resolutions and give an efficient algorithm that finds such a set optimally.

% -- Practical running time
The method for discovering topological domains that we have introduced is practical for existing datasets.  Our implementation is able to compute the consensus set of domains for the human fibroblast cell line and extract the consensus set in under 40 minutes when run on a personal computer with 2.3GHz Intel Core i5 processor and 8Gb of RAM.

Our method is particularly appealing in that it requires only a single user-specified parameter $\gamma_{\text{max}}$. It uses a score function that encodes the quality of putative domains in an intuitive manner based on their local density of interactions.  Variations of the scoring function in~(\ref{quality}), for example, by median centering rather than mean centering, can be explored to test the robustness of the enrichments described here. For our experiments, the parameter $\gamma_{\max}$ was set based on the maximum domain sizes observed in Dixon et. al's experiments so that we could easily compare our domains to theirs.  This parameter can also be set intrinsically from properties of the Hi-C interaction matrices.  For example, we observe similar enrichments in both human and mouse when we set $\gamma_{\max}$ to be the smallest $\gamma \in \Gamma$ such that the median domain size is $>$80kbp (two consecutive Hi-C fragments at a resolution of 40kbp). This is a reasonable assumption since domains consisting of just one or two fragments do not capture higher-order spatial relationships (e.g. triad closure) and interaction frequencies between adjacent fragments are likely large by chance~\cite{LiebAid2009}.  We also compared the fraction of the genome covered by domains identified by Dixon et al. vs. the domains obtained from our method at various resolutions.  Dixon et al.'s domains cover 85\% of the genome while our sets tend to cover less of the genome ($\approx$ 65\% for a resolution which results in the same number of domains as those of Dixon et al.).  The fact that our domain boundaries are more enriched for CTCF sites indicates that our smaller, more dense domains may be more desirable from the perspective of genome function.

The dense, functionally-enriched domains discovered by our algorithm provide strong evidence that alternative chromatin domains exist and that a single length scale is insufficient to capture the hierarchical and overlapping domain structure visible in heat maps of 3C interaction matrices. Our method explicitly incorporates the desirable properties of domain density and persistence across scales into objectives that maximize each and uncovers a new view of domain organization in mammalian genomes that warrants further investigation.

\section{Acknowledgments}

This work has been partially funded by National Science Foundation (CCF-1256087, CCF-1053918, and EF-0849899) and National Institutes of Health (1R21AI085376). C.K. received support as an Alfred P. Sloan Research Fellow. D.F. a predoctoral trainee supported by NIH T32 training grant T32 EB009403 as part of the HHMI-NIBIB Interfaces Initiative.

\nocite*
\bibliographystyle{plain}
\bibliography{coredomain}

\end{document}